\documentclass{v17dm}

\usepackage{amsmath}
\usepackage{amsfonts}
\usepackage{amssymb}
\usepackage{slashed, cancel}
\usepackage{geometry}

\newcommand{\GeV}{{\rm \,GeV}}
\newcommand{\TeV}{{\rm \,TeV}}
\newcommand{\ifb}{\,{\rm fb}^{-1}}

\def\eg{{\it e.g.}}

\def\pT{p_{\rm T}}
\def\mX{m_{\chi}}
\def\gX{g_{\chi}}
\newcommand{\met}{\slashed{E}_T}

\newcommand{\mZp}{m_{Z'}}
\newcommand{\Mmed}{M_{\rm med}}

\newcommand{\Photo}{\includegraphics[height=35mm]{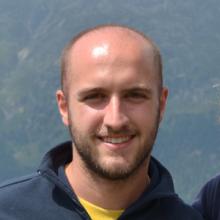}}

\begin{document}
\vspace*{4cm}
\title{Non-SUSY WIMPS: simplified models and Dark Matter @ LHC}

\author{Enrico Morgante}

\address{DESY, Notkestra{\ss}e 85, D-22607 Hamburg, Germany}

\begin{flushright} 
DESY 17-171
\end{flushright}

\maketitle

\abstracts{In this talk I discuss the construction of Simplified Models for Dark Matter searches at the LHC. After reviewing the phylosophy and some simple example, I turn the attention to the aspect of the theoretical consistency of these models.}


\section{Introduction}
\label{sec:IntroSimpMods}

The most studied solutions to the ``naturalness'' problem of the Standard Model rely on the existence of some new physical phenomena at the TeV energy scale. Essentially all of the naturalness-inspired scenarios can accomodate the presence of a good Dark Matter candidate: a neutral and very long-lived particle that was copiously produced in the early universe and then lost thermal contact with the SM (if it ever occourred) leaving a relic density $\Omega_{\rm DM}\sim 0.26$ of cold particles.
The LHC is a perfectly suited machine to look for this kind of particles, and current bounds from Atlas and CMS complement those from direct and indirect searches.

Given the plethora of particle physics model beyond the SM providing a WIMP candidate, it is highly desirable to study the signatures of this DM candidate in a model-independent way.
In this sense, the simplest approach is that of relying on a set of non renormalizable operators, that parametrize the interaction of the DM particle with SM fields in terms of one effective scale $\Lambda$, in addition to the DM mass $\mX$\cite{Goodman:2010ku}.
This approach was widely used for early Run-I analyses, and has the great advantage of giving bounds that are as model-independent as possible: if we consider a single effective operator and we extrapolate the EFT to high energies, the potential number of WIMP models is reduced down to a relatively small basis set.
Since direct and indirect detection of WIMPs, as well as WIMP production at the LHC, all require an interaction of the WIMPs with the SM particles, and such an interaction may be generated by the same operator, the EFT approach has the additional advantage of facilitating the analysis of the correlations between the various kinds  of experiments.

Despite its advantages, the EFT description has the drawback of an intrinsic energy limitation. At energies larger than $\Lambda$, which can be regarded as the cutoff of the effective theory, predictions become unreliable because of spurious effects such as violation of perturbative unitarity.
In general, if the EFT is regarded as the low energy limit of a theory with a heavy mediator of mass $M$, the cutoff is obtained as $\Lambda^2\sim M^2/g^2$, where $g$ is some combination of the coupling constants, and the theory is valid up to $p^2\lesssim M^2 \sim \Lambda^2$, where $p^2$ is the momentum exchanged in the process, and in the last passage we have assumed $g\sim 1$.\footnote{The actual expression of $p^2$ as a function of the external particles' momenta depends on the Feynman diagrams that enter the process, and it is a model dependent quantity.}
While this constraint is normally satisfied in Direct and Indirect probes, at the LHC the typical value of the momentum exchange is larger than the values of $\Lambda$ that can be excluded within the EFT framework, making the na\"ive EFT bounds unrelaiable except for values of the couplings close to the perturbative bound $g\lesssim 4\pi$~\cite{Busoni:2013lha,Busoni:2014sya,Busoni:2014haa}. If smaller values are assumed, a recasting procedure must be adopted to rederive the bounds considering only a fraction of the events in the simulation that correspond to those which fullfil the requirement on the momentum~\cite{Busoni:2014sya,Racco:2015dxa}.

In more recent years, the LHC community has turned its attention to the toolkit of simplified models.
The idea of simplified models was firstly adopted in the context of Supersymmetry searches, as a way to grasp the most relevant features of different SUSY models which have similar signatures at colliders.
Such models are characterized by the most important state mediating the interaction of the DM particle with the SM, as well as the DM particle itself.
Including the effect of the mediator's propagator allows to avoid the energy limitation of the EFT, and the simplified models are able to describe correctly the full kinematics of DM production at the LHC, at the price of an increased number of parameters.
The effective scale $\Lambda$ is traded in for the mass of the mediator and a handful of coupling constants, which poses important questions about the best way to constrain the parameter space and to present results.
An additional advantage of simplified models over the EFT approach is that they allow to exploit the complementarity between different LHC searches, such as searches for narrow resonances in the di-jet channel or di-jet + MET searches.

In this contribution we are first going to describe the construction of DM simplified models from a bottom-up approach, following the way these models were developed by the community.
Secondly, we are going to highlight some of the theoretical issues that are present in the most na\"ive construction, that are leading to a new generation of a ``less simplified'' models built on more stable theoretical grounds.
The first part of this discussion will follow the presentation of the early white papers on the subject~\cite{Abdallah:2014hon,Malik:2014ggr,Abdallah:2015ter}, as well as the final report of the \emph{ATLAS-CMS DM forum}~\cite{Abercrombie:2015wmb}, that gathered together more than 100 people with the goal of defining recommendations for well-defined and exhaustive benchmark models and encouraging communication between experimentalists and theorists working on collider physics (and possibly other probes).
The discussion will be based on the presentation of~\cite{Tesi-Springer}.
For additional reference on the subject of simplified models and DM searches at the LHC in general we recommend the reading of~\cite{DeSimone:2016fbz,Kahlhoefer:2017dnp}.

\section{Phylosophy of simplified DM models}

As in the case of the EFT, the idea beyond simplified models is to provide a good representation of possibly all realistic WIMP scenarios within the energy reach of the LHC, restricting to the smallest possible set of benchmark models, each with the minimal number of free parameters.
Simplified models should be complete enough to give an accurate description of the physics at the scale probed by colliders, but at the same time they must have a limited number of new states and parameters.
More over, they should satisfy all constraints posed by low-$p_{\rm T}$ analysis, such as those coming from flavour physics.
A simple recipe for simplified model building is:
\begin{itemize}
\item
The Lagrangian must contain a stable DM candidate and a mediator that couples it to the SM.
All additional states should be decoupled if not necessary for the consistency of the model itself.
\item
The Lagrangian should contain all renormalizable terms consistent with Lorentz invariance, gauge symmetry and DM stability.
\item
{\it Ad hoc} simplifications may be achieved by setting some parameters to zero or taking some of them to be equal, but this should be implemented in such a way that the phenomenology is not totally altered, in order not to prejudice the credibility of the constraints on the model itself.
\item
Interactions that violate the accidental global symmetries of the SM model (both exact and approximate) must be handled with great care.
Indeed, constraints on processes that violate these symmetries are typically very strong, and may overcome those coming from DM searches or even rule out all of the interesting parameter space of the simplified model.
For this reason, lepton and baryon number conservation is typically assumed, together with Minimal Flavour Violation (MFV).
Even with this assumption, there are cases in which constraints from flavour physics may be stronger than those coming from mono-X searches \cite{Dolan:2014ska} (see also \cite{Agrawal:2014aoa} for a discussion of a non-minimally flavour violating dark sector).
\end{itemize}

Most simplified models of interest may be understood as the limit of a more general new-physics scenario, where all new states but a few are integrated out because they have a mass larger than the energy scale reachable at the LHC or because they have no role in DM interactions with the SM.
Similarly, in the limit where the mass of the mediator is very large, the EFT framework may be recovered by integrating out the mediator.
On the contrary, there are new physics models which can not be recast in terms of simplified models, typically because more than just one operators are active at the same time, and possibly interfere with each other.

It should be noticed that the correspondence between simplified models and EFT is not one to one.
Different simplified models may give rise to the same effective operator after some Fierz rotation, as pointed out in~\cite{Racco:2015dxa} with the example of a $Z'$ model and of a SUSY-inspired model with coloured scalar mediators in the t-channel.

From the point of view of LHC searches, the enlarged physical spectrum and parameter space of full new physics theories with respect to simplified models, and of the latter with respect to the EFT, lead to a greater variety of possible search channels.
While within the EFT approach the mono-X searches give the best sensitivity, simplified models of DM can be constrained also with multi-jet + MET searches, with di-jet resonance searches and others, depending on the degree of sophistication of the model.
To summarize, on the good side the additional degrees of freedom in going from the EFT to simplified DM models and to full theories  allow to put limits on the DM properties by exploiting new search channels and the complementarity with other experimental searches; on the bad side, it involves more model dependence and requires care in the choice of the parameters and in the presentation of results.

\section{Simple construction}

The first set of simplified models we want to consider is the one where the DM interacts with quarks through the exchange of a mediator in the $s$-channel.
Assuming the DM particle $\chi$ to be a fermion (either Dirac or Majorana), and assuming CP-conservation, the Lagrangian of our models are
\begin{eqnarray}
\mathcal{L}_{\rm S}		& \supset &	-\frac{1}{2}M_{\rm med}^2S^2-y_{\chi}S\bar{\chi}\chi-y_q^{ij}S\bar{q}_iq_j+{\rm h.c.}\,, \nonumber \\
\mathcal{L}_{\rm S'}	& \supset &	-\frac{1}{2}M_{\rm med}^2S^{\prime2}-y^\prime_\chi S^\prime\bar\chi\gamma_5\chi-y_q^{\prime ij}S\bar{q}_i\gamma_5q_j+{\rm h.c.}\,, \nonumber \\
\mathcal{L}_{\rm V}		& \supset &	\frac{1}{2}M_{\rm med}^2V_\mu V^\mu-g_\chi V_\mu\bar\chi\gamma^\mu\chi-g_q^{ij}V_\mu\bar q_i\gamma^\mu q_j\,, \nonumber \\
\mathcal{L}_{\rm V'}	& \supset &	\frac{1}{2}M_{\rm med}^2V^\prime_\mu V^{\prime\mu}-g^{\prime}_\chi V^\prime_\mu\bar\chi\gamma^\mu\gamma_5\chi-g_q^{\prime ij}V^\prime_\mu\bar q_i\gamma^\mu\gamma_5q_j\,.
\label{eq:Ssimp}
\end{eqnarray}
where $S,S',V,V'$ stand for a scalar, a pseudo-scalar, a vector or an axial-vector mediator respectively, $q=u,d$ and $i,j=1,2,3$ are flavor indices.

As concerns the mediator couplings to quarks, the existence of off-diagonal coupling is tightly constrained by various FCNC processes.
For this reason, a good choice is to force the couplings to be diagonal: $g_q^{ij}=g_q^i\delta^{ij}$.
As a further simplification, one could fix the couplings to be flavour blind
\begin{equation}
g_d^i=g_u^i\equiv g_q \quad {\rm for} \quad i=1,2,3\, ,
\end{equation}
or take the coupling to the third generation to be stronger than the others.
Common benchmarks for the vector (V) ad axial vector (A) models are~\cite{Boveia:2016mrp}:
\begin{equation}
V:\,\left\{
\begin{array}{l}
\gX=1, g_q=0.25, g_\ell=0 \\
\gX=1, g_q=0.1, g_\ell=0.01 \\
\end{array}\right.
\quad
A:\,\left\{
\begin{array}{l}
\gX=1, g_q=0.25, g_\ell=0 \\
\gX=1, g_q=0.1, g_\ell=0.1 \\
\end{array}\right.
\end{equation}

It should be noticed that the scalar and pseudo-scalar models of Eq.~\ref{eq:Ssimp} are not gauge invariant.
This may lead to spurious results in processes where a $W/Z$ boson is emitted, but results of jets + MET searches are expected to be only mildly affected by this issue \cite{Boveia:2016mrp}.
Moreover, in the axial vector model perturbative unitarity is violated in a large portion of parameter space \cite{Kahlhoefer:2015bea}, and the indication of where the violation happens should be clearly shown when presenting constraints on this model.
We will return to these issues in section~\ref{sec:simpmodsdiscussion}.

Constraints on $s$-channel simplified models have been obtained by numerous groups, in particular for exchange of a vector mediator (scalar mediators are more problematic, see the discussion in section~\ref{sec:simpmodsdiscussion}).
In the mono-jet channel, the analysis in \cite{Abdallah:2014hon} shows that, for $\Mmed \lesssim 2 \mX$, the LHC at $14\TeV$ with $300\ifb$ is sensitive to $\mathcal{O}(1)$ couplings only for $\mX\lesssim\mathcal{O}(100\GeV)$, while for $\mX\sim1\TeV$ it is sensitive only to couplings of order $\gX\cdot g_q \gtrsim 10$.
Mono-jet and mono-X constraints are discussed in details in~\cite{Jacques:2015zha,Brennan:2016xjh}, in which a rescaling procedure is proposed in order to set limits on the model's couplings in the plane $m_{\rm DM}-M_{\rm med}$ without the need of a three-dimensional scanning.
In~\cite{Chala:2015ama,Duerr:2016tmh} mono-jet searches are compared to dijet searches (see also~\cite{Fairbairn:2016iuf}), direct detection limits, dark matter overproduction in the early universe and constraints from perturbative unitarity.

A very interesting phenomenology arises in the case where the couplings to third generation quarks is larger than the couplings to the first two.
This may happen for example in models in which a scalar mediator is exchanged and MFV is assumed.
The Yukawa couplings of the mediator are then proportional to the fermions' mass, resulting in an enhanced coupling to $b$ and $t$ quarks.
Because of the peculiar signature of events with these quarks in the final state, very strong constraints on these models come from searches for one or two $b$-tagged jet~+~MET and $t\bar t$~+~MET (see \cite{Lin:2013sca} for an early proposal within the EFT framework and \cite{Buckley:2014fba,Abercrombie:2015wmb} for a discussion in terms of simplified models).

An interesting possibility is that the scalar mediator of the DM-SM interaction is the Higgs boson itself, as it happens in the ``Higgs portal'' models (see {\it e.g.}~\cite{Duerr:2015aka}).
Many manifestations of Higgs portal models would lead to a reduction or suppression of the Higgs boson couplings to SM particles, in favor of its interactions with new particles~\cite{Englert:2011yb}.
Precision measurements of the Higgs couplings that can be undertaken in future LHC phases and future accelerators can further constrain Higgs portal models~\cite{Englert:2014uua}.
Alternatively, the Higgs' coupling to DM can be constrained by measurements of the Higgs partial width to invisible particles.
Current ATLAS and CMS limits on invisible Higgs decay at the 95\% C.L. are around 70\%; they are expected to decrease to 20-30\% by the end of the $300\,\ifb$ LHC run~\cite{ATL-PHYS-PUB-2013-015}.

Another interesting possibility is that of a coloured fermionic mediator with an interaction vertex between quarks and the WIMP resulting in a $t$-channel exchange, as with squark in supersymmetric models.
\begin{equation}
\mathcal L = \mathcal{L}_{SM} + g_{M}\sum_{i} \left(\bar{Q}^i_L \widetilde{Q}_L^i  +  \bar{u}^i_R \tilde{u}^i_R + \bar{d}^i_R \tilde{d}^i_R \right)\chi + \text{mass terms} +c.c. 
\label{eq:lagrangian}
\end{equation}
where $Q_{L}^i, u_R^i, d_R^i$ are the usual SM quarks, $\widetilde{Q}_L^i, \widetilde{u}_R^i, \widetilde{d}_R^i $ correspond to the respective squarks, and $i$ represents a flavour index.
Unlike the usual case in Superysmmetry, here the WIMP $\chi$ can be taken to be either Dirac or Majorana fermion.
This model is extensively analysed in \cite{Papucci:2014iwa}, and a comparison with its effective operator limit is performed.

Two interesting features of this model are worth listing, that makes it qualitatively different from its low energy EFT limit.
Firstly, being the squarks coloured, gluons may be emitted not only as initial state radiation but also from the mediator itself.
This process is suppressed in the EFT limit by two powers of $\Mmed$, and this make a large qualitative difference in the kinematic distribution within the simplified model and the corresponding operator.
Secondly, when the mediator is light enough, its pair production becomes kinematically accessible, and an event like $p\,p \to \widetilde{q}\,\widetilde{q} \to q\chi \,q\chi$ leads to a di-jet + MET signature (or in general jets + MET, when additional jet radiation is taken into account).
It was shown in \cite{Papucci:2014iwa} that, on a large portion of parameter space, this kind of signature with two high-$\pT$ jets leads to constraints stronger then the mono-jet one, even when the effect of additional sub-dominant jets is taken into account.
More over, the effect of off-shell $\widetilde q$ production is important, and the effect of a finite width should be taken into account.

\section{Simplified models - a critical look}
\label{sec:simpmodsdiscussion}

The list of simplified models presented in the previous section was constructed keeping in mind the EFT approach and its limitations. In this sense, simplified models can be viewed as an improvement of effective operators, where the effective scale $\Lambda^4$ is replaced by a propagator's denominator $(p^2-M^2)^2+\Gamma^2 M^2$ in order to avoid energy limitations and exploit resonance enhancement in the production cross section. In a bottom-up approach, this is a small step above.
As soon as we start looking at it more carefully, anyway, the situation turns out not to be so simple.

The first problem lies in the fact that the simplified models of Eq.~\ref{eq:Ssimp} are not invariant under the full SM gauge group ${\rm SU}(3)_c\times{\rm SU}(2)_L\times{\rm U}(1)_Y$ but only under the unbroken subgroup ${\rm SU}(3)_c\times{\rm U}(1)_{\rm e.m.}$. This is true for models with a scalar mediator (in which Yukawa couplings break ${\rm SU}(2)_L$), and for models with a vector mediator in which the couplings to up and down quarks are different (so that the mediator does not couple to the left handed quark doublet but to its two components separately).
Violation of the electroweak gauge symmetry can lead to spuriously enhanced cross section for DM production with the initial state radiation of a $W$ boson~\cite{Bell:2015sza,Bell:2015rdw}. This problem does not only affect the mono-$W$ searches: the $W$ can indeed decay hadronically, enhancing the signal in the mono-jet search. For example, in the case of a vector mediator with opposite sign couplings to up and down quarks, this process dominates the mono-jet cross section for $\met>400\GeV$~\cite{Haisch:2016usn}.
This means that, even when restricting to a particular MET search, constraints descending from the internal consistency of the model can not be neglected.

There are of course ways to cure the models by adding new particles or new interactions. Again referring to the case of a vector mediator, different couplings of the up and down quarks can be made compatible with perturbative unitarity if an appropriate vertex $WWZ'$ is added (where $Z'$ is the new vector mediator), in similarity to what happens for the $Z$ boson in the SM.
The situation is more complex in the case of the scalar mediator. There gauge invariance can not be simply restored with a choice of the couplings. The reason is that a singlet $S$ can couple to DM but not to quarks, while if $S$ is a doublet it can couple only to quarks, but not to DM. A possible solution is to add a mixing of $S$ with the Higgs boson, via a quadrilinear term $H^\dagger H S^2$. In this case, constraints on the Higgs width to invisible particles and direct detection force the mixing angle $\varepsilon$ (and therefore the coupling $g_q\sim\sin\varepsilon$) to be very small, making LHC constraints weak.
In turn, this may be overcome by adding to the model a second Higgs doublet and letting $S$ mixing with it, but then again the phenomenology is altered (see\eg~\cite{Goncalves:2016iyg,Bell:2016ekl}).

A second issue is again related to perturbative unitarity.
Let us consider a model with a spin-1 mediator with both vectorial and axial couplings to DM and to quarks:
\begin{equation}
\label{eq:L_VA}
\mathcal{L} =  - \sum_{f = q,l,\nu} Z'^\mu \, \bar{f} \left[ g_{f}^V \gamma_\mu + g_f^A \gamma_\mu \gamma^5 \right] f - Z'^\mu \, \bar{\psi} \left[ g_\text{DM}^V \gamma_\mu + g_\text{DM}^A \gamma_\mu \gamma^5 \right] \psi \; .
\end{equation}
Once applied to the elastic scattering of fermions (both SM fermions or DM) the perturbative unitarity bound on this model reads $m_f \lesssim \mZp/(\sqrt2 g_f^A)$~\cite{Kahlhoefer:2015bea}, where $f$ may stand for both a SM fermion or the DM particle.
In a similar way, perturbative unitarity is violated in the process of 2 fermions annihilation into $Z'Z'$, which is important for the calculation of the relic density.
In order to restore unitarity new physics has to be invoked. In particular, what violates unitarity is the longitudinal mode of the $Z'$ boson, therefore the addition to the model of a scalar particle that give rise to its mass via Higgs mechanism would serve the purpose.
In this case, the condition on the mass of the $Z'$ would read
\begin{equation}
\sqrt{\pi}\frac{\mZp}{g_{\rm DM}^A} \geq \max [m_s, \sqrt{2}m_{\rm DM}] \,,
\end{equation}
where $m_s$ is the mass of the new scalar.
Notice that, in any case, the problem of unitarity affects only the axial coupling of the $Z'$, while the vector coupling $g_f^V$ is not affected by these constraints (which indeed become trivial in the limit $g_f^A\to 0$).

In the example of a $Z'$ gauge mediator, not only an additional Higgs boson $s$ is necessary to give mass to the $Z'$ and to cure the unitarity issues, but also the SM Higgs boson must be charged under the new gauge group for the Yukawa couplings to be gauge invariant.
As discussed in~\cite{Jacques:2016dqz}, this leads to the necessary presence of a $Z'Zh$ vertex.
While one can, to a certain extent, ignore these complications at the LHC, this is not possible for indirect searches since tree level annihilation into $Zh$~\cite{Jacques:2016dqz} and $Zs$~\cite{Bell:2016fqf} are easily the dominant channels, together with loop annihilation into EW gauge bosons. More over, leptonic annihilation channels $\ell\bar\ell$ are at least as important as $q\bar{q}$ ones, and the coupling of the mediator to leptons can not be ignored as it is typically done for LHC searches. The same is true for the calculation of the relic density.

We would like to spend a last comment on the issue of gauge anomalies. If the interaction of DM with SM fermions is due to an extended gauge symmetry, in order for the theory to be consistent at the quantum level the charge assignment under the new gauge group has to be decided in such a way that gauge anomalies are avoided.
Alternatively, additional heavy fermions may be added to the model. If they are heavy enough, their impact on LHC searches is negligible, and the details of this part of the dark sector can be ignored. Unfortunately this is not the case when we want to compare LHC results with those from indirect searches: in the latter, indeed, loop annihilation channels are relevant, but these can not computed including SM fermions alone because divergences do not cancel, and the full knowledge of the dark sector is needed~\cite{Jacques:2016dqz} (see also~\cite{Ellis:2017tkh,Ismail:2017ulg}).

\section{Conclusions}

In the way they were originally introduced, simplified models were meant to be a small step beyond the EFT approach, in such a way to avoid their intrinsic energy limitation and to exploit resonant production of the mediator in order to improve the constraining power of LHC searches.
It was successively realised that, in order not to deal with unphysical results, the vanilla picture had to be supplied with additional constraints, couplings and states, in a kind of second order improvement. On the one hand, the typical consequence is that the strong LHC constraints on the dark sector do not come from DM production processes (as in mono-X searches) but from other observables (di-jet and di-lepton resonances, mixing with $Z$ boson and electroweak precision tests, Higgs width to invisibles, perturbative unitarity ${\it etc.}$~\cite{Duerr:2016tmh}).
This comes with no surprise, since the high energy reach of the LHC consents to explore a large variety of phenomena above the weak scale, without restricting to the lightest stable state of this new physics sector. This is quite the opposite with respect to what happens with direct and indirect searches, which are intrisically limited to constrain the properties of the DM particle.
On the other hand, simplified models can not (or only partially) be viewed as an exhaustive toolbox to constraint all possible WIMP scenarios at once.
For this reason, it is of extreme importance that the LHC collaborations publish their results on simple, search-specific, models in such a way that they are recastable for any other model (as it is for cut-and-count analyses). In turn, theoreticians should keep working in close contact with experimentalists in order to maximise the utility of the simplified models toolkit.
Finally, the use of (truncated) EFT should not be disregarded, since this is the most model independent approach and it is economical from the point of view of the computational effort because of the reduced dimensionality of its parameter space (and therefore of parameter scannings).

\section*{References}

\bibstyle{unsrt}
\bibliography{lit}


\end{document}